# Quantum oscillation and topology change of the uncondensed Landau Fermi surface in superconducting CeCoIn$_5$

Sangyun Lee[1], Duk. Y. Kim[1], Andrew J. Woods[1], Priscila F. S. Rosa[1], E. D. Bauer[1],

Filip Ronning[1], Shi-Zeng Lin[1], R. Movshovich[1,*]

[1]Los Alamos National Laboratory, Los Alamos, New Mexico 87545, USA

[*] Corresponding author: R. M. (roman@lanl.gov)

**Abstract**

Metals typically have multiple Fermi surface sheets, and when they enter the superconducting state, some electrons on these sheets may remain uncondensed, or their superconducting pairs can be rapidly destroyed by a magnetic field. Detecting uncondensed electrons within the superconducting state provides key information about the underlying electronic structure; however, this task remains a significant experimental challenge. Here we demonstrate quantum oscillations from the uncondensed electrons in the heavy-fermion superconductor CeCoIn$_5$, observed through thermal conductivity measurements with a magnetic field rotating within the tetragonal *a-b* plane. We detect a fine structure in thermal conductivity, characterized by multiple small resonances (oscillations) in a rotating magnetic field. Remarkably, the phase of these resonances shifted by as much as $\pi$ for a field above 9.7 T where spin-density wave (SDW) order emerges and coexists with superconductivity. This phase shift is naturally explained by a change in the Berry phase of the uncondensed Fermi surface, driven by the Fermi surface reconstruction associated with the onset of SDW order. Our work unambiguously shows the existence of uncondensed electrons in the superconducting state of CeCoIn$_5$, thus resolving a longstanding debate on this issue.

**Introduction**

CeCoIn$_5$ is a tetragonal heavy-fermion superconductor with a $d_{x^2-y^2}$–wave superconducting order parameter, which hosts four nodes within the *a-b* plane. With a relatively high superconducting transition temperature ($T_c$) of 2.3 K at ambient pressure, it stands out as an ideal platform for investigating unconventional superconductivity (SC) in heavy-fermion materials [1-4]. The superconducting upper critical field ($H_{c2}$) reaches 11.5 T for fields applied within the *a-b* plane, whereas $H_{c2}$ = 5 T along the *c* axis. With a large Maki parameter of 4.5, $H_{c2}$ is primarily constrained by the Pauli mechanism [5,6], which makes CeCoIn$_5$ a strong candidate to host the Fulde-Farrell-Larkin-Ovchinnikov (FFLO) phase [7, 8]. Though in zero field CeCoIn$_5$ does not order magnetically, a Spin Density Wave (SDW) with an ordering wavevector $Q$ = (0.44, ±0.44, 0.5) along the two nodal direction of the $d_{x^2-y^2}$– wave superconducting order parameter is observed for an in-plane magnetic field above 9.7 T but below $H_{c2}$ [9-11]. The $Q$ direction can be switched sensitively by the in-plane magnetic field direction, and a third intertwined order (a secondary superconducting phase) has been suggested [12]. The SDW state coexists with SC, representing a rare instance of coupled magnetic and superconducting order.

CeCoIn$_5$ has multiple Fermi sheets according to both density functional theory calculations and observation of quantum oscillation [13-15]. The appearance of multiple gaps in the superconducting state is also observed experimentally [16]. In Ref. [17], it is reported that uncondensed electrons coexist with nodal quasiparticles associated with the *d*-wave order parameter from the measurement of the residual thermal conductivity at zero magnetic fields as the temperature is extrapolated to $T$ = 0 [18]. In this case, SC arises from the pairing of heavy electrons, and unpaired electrons reside in the light bands [17,19]. However, later thermal conductivity measurements down to 10 mK suggest the absence of unpaired electrons and point to multigap SC [16,20]. These two seemingly contradictory conclusions remain unsolved and the existence of unpaired electrons in CeCoIn$_5$ is an open question.

Probing intrinsic properties of superconductors—such as the pairing symmetry, low-energy quasiparticle behaviours, or nodal gap structure—can be challenging due to the fully or partially gapped nature of the superconducting state, expulsion of magnetic fields (Meissner effect), and complications arising from vortex dynamics, surface effects, or impurity scattering. Conventional electric transport measurement is no longer able to detect quasiparticle excitations in a superconducting sample because electrical transport is shunted by the superconducting condensate.

Since the superconducting condensate does not carry entropy, thermal conductivity becomes a unique tool exquisitely positioned to study quasiparticle excitations [21]. For example, the change in the density of states of Bogoliubov quasiparticles induced by the Volovik effect, which manifests as a field-induced increase in the quasiparticle density of states in nodal superconductors, can be probed by measuring thermal conductivity while varying the direction and/or strength of the magnetic field with respect to the heat current.

If SC order parameter has nodes, the Bogoliubov quasi-particle density of states, dominated by the nodes, can be easily observed by thermal conductivity measurements [22,23]. In the case of CeCoIn$_5$, a four-fold oscillation of thermal conductivity as a function of field angle, one of the signature of the *d*-wave superconductivity, was observed while rotating magnetic field within the *a-b* plane [1,5,24]. Similarly, the appearance of SDW in CeCoIn$_5$ has also been detected by thermal transport measurements due to simultaneous gapping of the Bogoliubov quasi-particles at the two opposite nodes of the SC order parameter [25]. Besides the Bogoliubov quasi-particles, thermal current is also carried by uncondensed electrons, the same as those in normal metals. Under a strong magnetic field, these uncondensed electrons form Landau orbitals and exhibit quantum oscillations when magnetic field strength or orientation is varied.

Generally, the distribution of the magnetic field in the superconducting phase is not uniform in space due to the presence of the Abrikosov vortex lattice. However, the field variation is small in extreme type-II superconductors when the field is much larger than $H_{c1}$. The Bogoliubov quasi-particles also respond to the magnetic field, but in a distinct way because they do not have a well-defined charge. Instead of forming Landau levels, the Bogoliubov quasi-particles form the Bloch band by coupling to the Abrikosov vortex lattice. [26-28] This allows us to distinguish between uncondensed electrons and Bogoliubov quasi-particles associated with the superconducting condensate.

Therefore, thermal conductivity measurements serve as a powerful tool for probing the detailed nature of the superconducting state, such as distinguishing Bogoliubov quasi-particles from uncondensed electrons. In this study, we applied thermal conductivity measurements under a magnetic field, with the magnetic field orientation systematically varied. Our results provide unambiguous experimental evidence of the existence of unpaired electrons from the quantum oscillation of thermal conductivity in the superconducting state. Additionally, we also observe a change in the Berry phase of these unpaired electrons across the field-induced SDW transition. Besides confirming the pairing symmetry of the superconducting order parameter, our work establishes magnetic field-dependent thermal conductivity as an important tool to uncover the Fermi surface topology of the uncondensed electrons inside the superconducting phase.

**Results and Discussion**

Fig. 1(a) presents the phase diagram of $CeCoIn_5$ and a schematic of an experimental setup, with the angle ($\theta$) between the rotating magnetic field ($H$) and the heat current ($J$). $CeCoIn_5$ is a $d_{x^2-y^2}$-wave SC with four nodes within the crystallographic $a$-$b$ plane. A needlelike single-crystal sample was prepared with a long axis along the [100] crystallographic direction, which coincides with superconducting antinodes. Heat current was applied along the *a*-axis ([100] direction), and thermal conductivity was measured with the standard steady-state method with one heater and two thermometers. The magnetic field was applied within the *a-b* plane at a fixed temperature of 90 mK and the crystal was rotated about the *c*-axis using an Attocube piezoelectric rotator [12,29], with $\theta = 0$ corresponding to the field parallel to the [100] direction.

SDW appears when the magnetic field exceeds 10 T, as depicted in the phase diagram of Fig. 1(b). We investigated CeCoIn$_5$'s normal quasi-particle density of states using thermal conductivity measurements in the region of the *Q*-phase (SDW) above 10 T as well as below 10 T in the regime of uniform *d*-wave SC. In Fig. 1(b), the red dashed line in the phase diagram represents where we measure the thermal conductivity.

Fig. 2(a) shows the measured thermal conductivity divided by temperature ($\kappa/T$) in a wide range of magnetic fields from 1 T to 11 T rotating around the *c*-axis from -100 ° to 100 °. At 1 T, a 4-fold cosine fit is plotted with the data. $\kappa/T$ roughly shows 4-fold behavior, which is also reported in previous studies [1]. In addition, there are two major interesting features. First, as the magnetic field increases, a large peak at $\theta = 0$ ° and two small minor peaks around $\theta = \pm 12$ ° evolve, as marked by the red and blue arrows in Fig. 2(a), respectively. Second, small resonances or step-like features are resolved in the region from about -30 ° to +30 °, as shown in Fig. 2 (b) and (c).

The peak at $\theta = 0$ ° steadily increases with the rising magnetic field, all the way to 11 T. As shown previously, this peak is due to the vortex core states, which contribute the most to thermal conductivity when the applied field is parallel to the heat current [24]. The two smaller peaks around ±12 °, marked by red arrows, are observed at intermediate magnetic fields, and reported on previously [24]. These peaks seem to have the same origin as the peaks at ±33° away from the direction of the heat current applied along the [110] node, observed earlier [5] for magnetic field rotated within the *a-b* plane. The peaks observed in the two experiments correspond to the same direction of the magnetic field within the crystal lattice, independent of the heat current direction. Therefore, these peaks do not originate from the quasiparticle scattering but instead reflect intrinsic features of CeCoIn$_5$, such as the details of the Fermi surface. [24]

The fine structure of thermal conductivity is revealed in Figs. 2(b) and (c), where small resonances in $\kappa/T$ with magnetic field rotating in the *a-b* plane are marked by black arrows. Fig. 2(b) presents data at 11 T, with black arrows indicating the peaks of the small resonances, and the black solid lines guide the eye. These small resonances are observed at the lower field as well, down to 7 T. To analyze those small resonances as a function of angles, we fit and subtract the following background:

$$f'(n, \theta) = \sum_{k=0}^{n} C_{2k} \cos(2k\theta),$$

which is represented by the red solid lines in Fig. 2(c). As shown in Fig. 2(a), the 4-fold model using $f'(n=2, \theta) = C_0+C_2\cos(2\theta)+C_4\cos(4\theta)$ (black solid line) is insufficient to accurately describe the background. By extending the value of $n$ up to 6 ($f'(n=6,\theta)$) over a smaller range of the angle of field orientation, we achieved a more accurate representation of the background, as shown in Fig. 2(c). This extension allows us to extract the small resonance as a function of angles, which is induced by the rotation of the magnetic field orientation in the *a-b* plane.

Fig. 3 shows the results after subtracting the backgrounds represented by the red lines from Fig. 2(c) for each magnetic field. After the subtraction, a clear resonance behaviour appears as a function of the field angle. The resonances are particularly clear in the range of angles from -25 ° to -15 °, shown in the shaded region of Fig. 3. Outside the shaded range, the resonances are not as clear, probably due to a very steep background that is difficult to remove accurately. As Figs. 3(a)-(e) show, these resonances are pronounced after subtraction of the backgrounds for the field from 11 to 7 T. The data show a dip at -18.2 ° for field $H = 11$ T and 10 T (red arrows). However, as the strength of the magnetic field decreases, the dip at -18.2 ° changes to a peak below 9.5 T (black arrows).

In Fig. 3(f), the positions of all peaks and dips collected in different magnetic fields are plotted, and peaks and dips are marked as open and solid symbols, respectively. The dashed lines guide the eyes and show how the peaks and dips change as the magnetic field increases for each angle. As the strength of the magnetic field increases, the dips (peaks) switch to peaks (dips) at around 10 T, where the SDW order onsets and coexists with the *d*-wave superconductor (SC+SDW). For example, at the dashed line located at about -25 º below 10 T, the dips are formed for each rotating magnetic field, but peaks are formed above 10 T in the SC+SDW phase.

There are two main experimental observations: (1) peaks/dips in the thermal conductivity at a certain field angle in the SC phase, and (2) the peaks and dips swap positions when the system enters the SC+SDW phase. Here we provide a physical interpretation of these two main observations. The resonance around $\theta = \pm 12º$ was observed previously and was assigned as a resonance of the normal Fermi surface [24]. Such resonance can be assigned to a peak in the density of state (DOS) for different field angles calculated using the Fermi surface obtained from density functional theory (DFT) [24]. The DFT results reveal multiple peaks as a function of the magnetic field orientation (see, e.g., Fig. 5 in Ref. [24]). We argue that the mechanism of the resonances observed in Fig. 3 is the same as that around $\theta = \pm 12º$, which is due to the uncondensed normal Fermi surface, for the following reasons. Firstly, all the peaks within the same state are independent of the magnetic field strength, which is characteristic of the normal Fermi surface. The Bogoliubov quasiparticles can also develop resonance-like behavior when coupled to the background vortex lattice [31]; however, their resonance position depends on the magnetic field strength because the vortex lattice spacing varies with field strength. Secondly, the peak positions in the SC state, i.e., around $\theta = 13.9º$, $\theta = 18.7º$, and $\theta = 22.8º$ in Fig. 3(f) have a one-to-one correspondence to the DOS peaks according to the DFT calculations [24]. These resonance angles depict a fine structure of the Fermi surface, as schematically illustrated in Fig. 4 (a). When the field angles correspond to the extremal Landau orbitals determined by the saddle point of the Fermi surface with respect to the field angle, resonance occurs, resulting in a peak in thermal conductivity.

Next, we examine the observed reversal of peak and dip positions in the thermal conductivity as the system transitions from the pure SC phase to the coexisting SC+SDW phase. This behaviour can be understood in terms of a change in the Berry phase associated with the Fermi surface, which determines the offset in the oscillation as a function of magnetic field [35]. The swap of the peak and dip positions across the SDW transition indicates a $\pi$ phase shift due to $\pi$ Berry phase change. When the SDW order develops, translational symmetry is broken, leading to the reconstruction of the band structure due to band folding. It is possible that the Fermi surface with $\pi$ Berry phase is gapped due to the hybridization caused by band folding, as illustrated in Fig. 4 (b). This band reconstruction modifies the Berry curvature distribution, effectively altering the $\pi$ Berry phase around the Fermi surface. Consequently, the thermal conductivity is expected to exhibit an inversion of peak and dip features, consistent with experimental observations [32,33,34]. We note that, while the experimental observations can be explained by a change in the Berry phase by approximately $\pi$, this phase shift is not necessarily quantized to exactly $\pi$. This is due to the breaking of time-reversal symmetry by the external magnetic field and the presence of the SDW order parameter. To describe quantitatively the experimental observations, it will be important to go beyond this simplified picture and consider more realistic models that incorporate material-specific band structures and include both SC and SDW instabilities.

**Conclusion**

In summary, thermal conductivity measurements with rotation of magnetic field orientation have emerged as a novel tool for probing quantum oscillations within the superconducting state. This approach can identify uncondensed Fermi surfaces, determine the extremal Landau orbitals, detect potential Bogoliubov quasi-particle density of states, and, crucially, capture changes in the Berry phase, reflecting the topological nature of the Fermi surface. In our study, we performed high-resolution thermal conductivity measurements on the heavy-fermion superconductor CeCoIn$_5$ under rotating magnetic fields, both within the Q-phase—where spin-density wave (SDW) order coexists with superconductivity (SC)—and in the adjacent, uniform SC phase. In both regions, we observed small resonances in the thermal conductivity as a function of field angle. Notably, the phase of these resonances exhibited a sharp $\pi$ shift at the phase boundary between the two states, indicating a significant topological change in the Landau Fermi surface induced by the SDW order. This topological shift likely affects two of the four nodes of the $d_{x^2-y^2}$–wave superconducting order parameter. The resulting modification provides unambiguous evidence for the presence of uncondensed (normal) electrons in CeCoIn$_5$ and offering critical insights into the interplay between magnetism and superconductivity.

## Acknowledgments

Work at Los Alamos was performed under the auspices of the US Department of Energy, Office of Science, Division of Materials Science and Engineering. Theoretical work (SZL) was supported by DOE via LDRD program at LANL and in part, by the Center for Integrated Nanotechnologies, an Office of Science User Facility operated for the U.S. DOE Office of Science, under user proposals #2018BU0010 and #2018BU0083.

## Additional information

Correspondence and requests for materials should be addressed to RM (roman@lanl.gov).

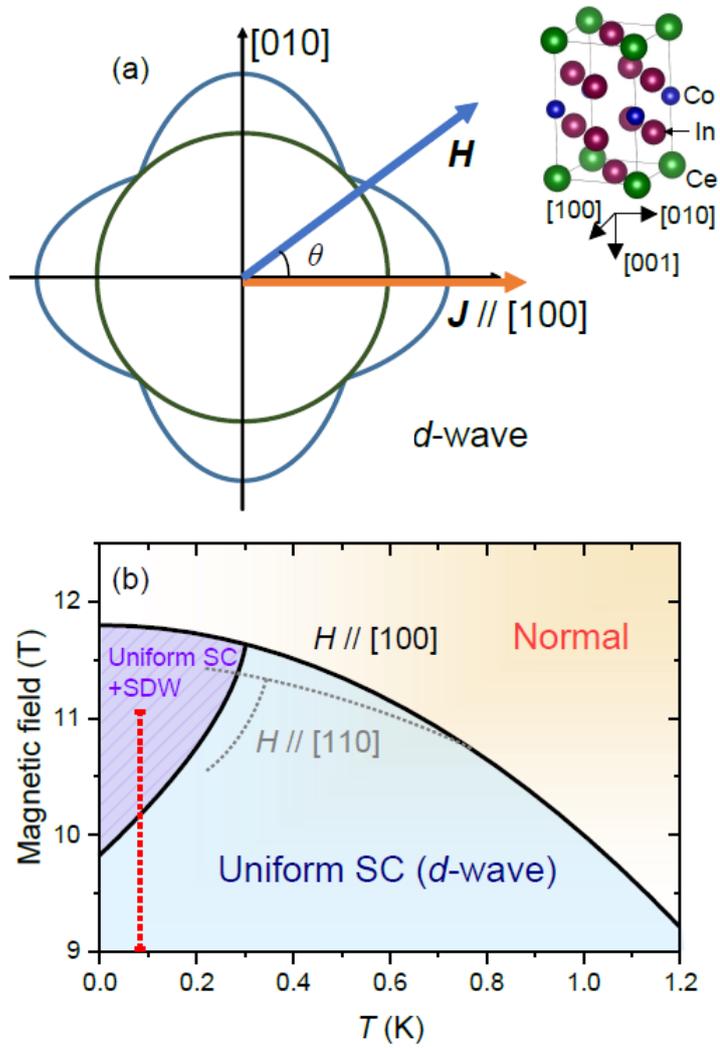

Figure 1. (a) Schematic diagram of thermal conductivity $\kappa$ measurement on $d_{x^2-y^2}$-wave superconductor CoCoIn$_5$. The heat current $J$ is applied along [100]. The angle $\theta$ is between the direction of heat current and magnetic field $H$. Inset indicates the crystal structure of CeCoIn$_5$. (b) Schematic phase diagram, showing the SDW phase at low temperature near upper critical field.

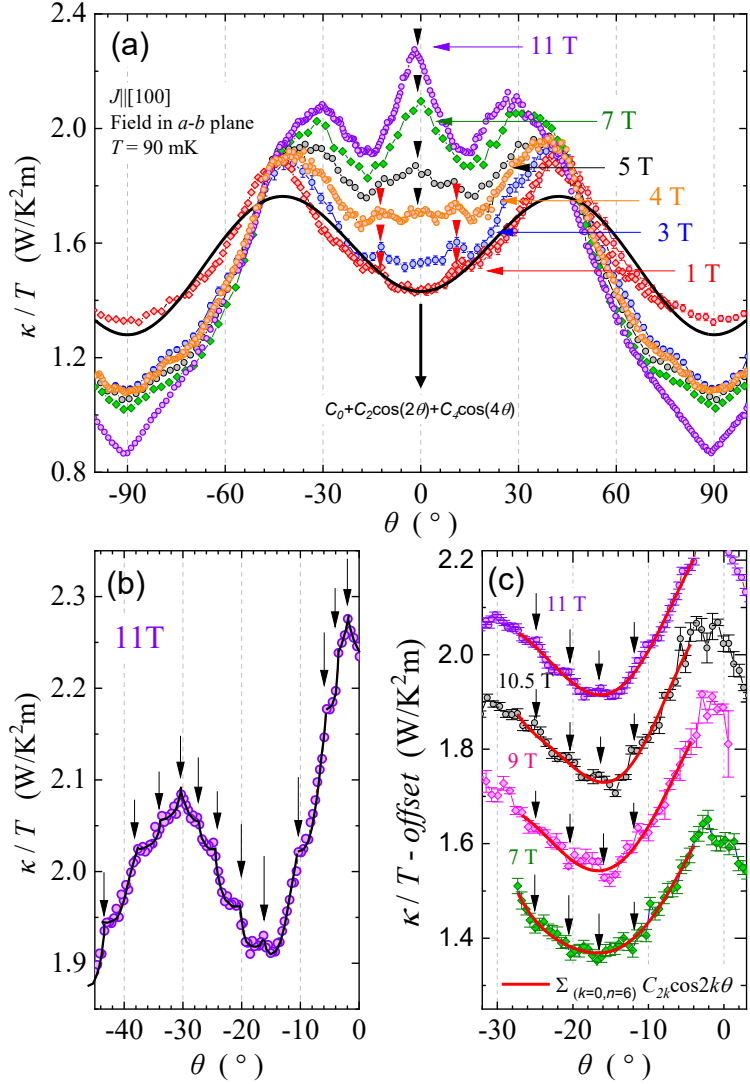

Figure 2. (a) The thermal conductivity of CeCoIn$_5$ divided by temperature $\kappa/T$ as a function of the angle for $J//[100]$ from 1 to 11 T. The Solid line represents a four-fold fit to the equation: $f(\theta) = C_0+C_2\cos(2\theta)+C_4\cos(4\theta)$. Near ±13º, small peaks are pronounced, and at 0º, the peak is evolved with increasing magnetic field. (b) For 11 T, a small resonance is pronounced in $\kappa/T$ as a function of the angle. Black arrows represent these resonance behaviours at 11 T. (c) Under magnetic field, small resonances appear. Red lines represent the fit to the equation $f'(\theta) = \Sigma(k=0, n=6)C_{2k}\cos(2k\theta)$. $\kappa/T$ as a function of the angle for different fields are rigidly shifted for clarity. Black arrows indicate positions of the resonances for $H > 10$ T, and anti-resonances for $H < 10$ T.

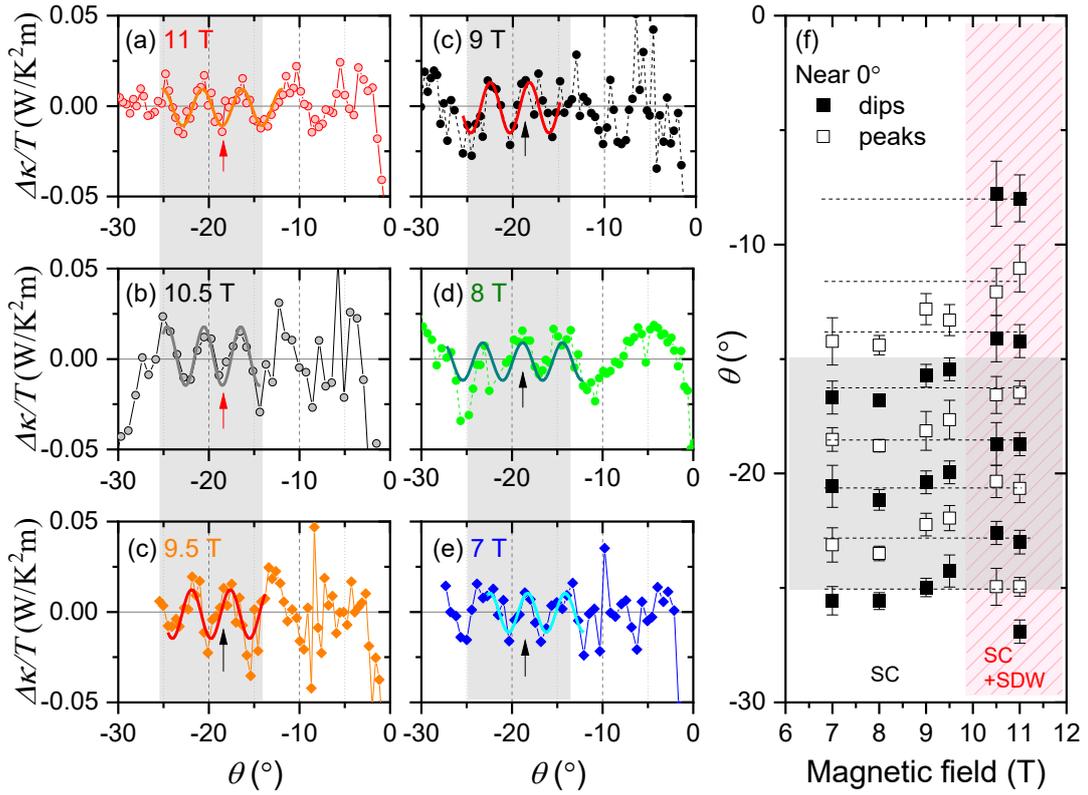

Figure 3. (a-e) The best-fitting function of $f'(\theta)$, represented by a red line in Fig. 2(c), is subtracted from the $\kappa/T$ as a function of the angle. Red and black arrows indicate the dip and the peak at -18.2°. (f) The dips and the peaks are marked as a function of the magnetic field for each angle. The grey shadow indicates where the dips and peaks are represented as a function of angles. The Red dashed shadow represents the region where SDW is observed with SC in the high-field low-temperature (HFLT) phase.

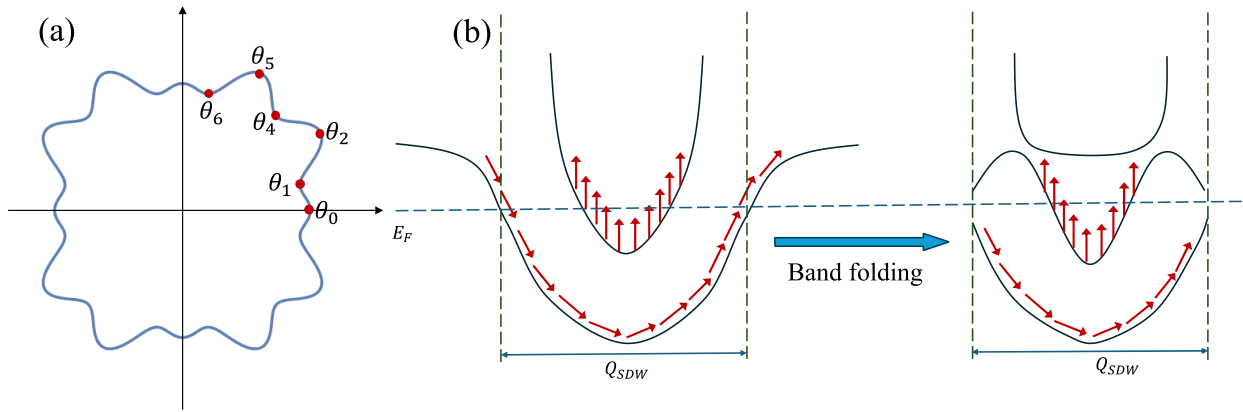

**Figure 4.** (a) A schematic Fermi surface with many saddle points highlighted by red dots. The Landau orbitals connecting antipodal saddle points generated by in-plane magnetic field are the extremal Landau orbitals. (b) An example illustrates the change of the Berry phase at the Fermi surface due to band folding. Red arrows depict the Berry phase of each band. The Fermi surface of the lower band has $\pi$ Berry phase, while the Berry phase for the upper band is 0. Upon band folding at the SDW wavevector $Q_{SDW}$, the Fermi surface of the lower band is gapped out, while the upper band remains gapless, resulting in a $\pi$ change in the Berry phase.